# DurableFS: A File System for Persistent Memory

Chandan Kalita, IIT Guwahati
chandan.kalita@iitg.ernet.in

Gautam Barua, IIIT Guwahati
gb@iiitg.ac.in

Priya Sehgal, Netapp India Pvt. Ltd.
Priya.Sehgal@netapp.com

## Abstract

With the availability of hybrid DRAM – NVRAM memory on the memory bus of CPUs, a number of file systems on NVRAM have been designed and implemented. In this paper we present the design and implementation of a file system on NVRAM called DurableFS, which provides atomicity and durability of file operations to applications. Due to the byte level random accessibility of memory, it is possible to provide these guarantees without much overhead. We use standard techniques like copy on write for data, and a redo log for metadata changes to build an efficient file system which provides durability and atomicity guarantees at the time a file is closed. Benchmarks on the implementation shows that there is only a 7% degradation in performance due to providing these guarantees.

## 1. Introduction

With the availability of hybrid DRAM – NVRAM memory on the memory bus of CPUs, a number of file systems on NVRAM (which we also refer to as Persistent Memory (PM)) have been designed and implemented [3,4, 12, 13]. On the one hand, the fast storage randomly accessible at the byte level, provides opportunities for new file system designs, but on the other hand, the presence of a cache hierarchy on the path to the NVRAM and a lack of feedback when data actually reaches NVRAM, poses challenges in providing guarantees on the durability of operations. In this paper, we present the design and implementation of a file system for NVRAM called DurableFS. Previous file systems for NVRAM are POSIX compliant, with durability semantics as is present in UNIX and Linux disk file systems. The main goal of durability in these file systems is to enable a consistent state of the file system to be available after a crash. This involves logging of only metadata changes. Durability of individual file data operations is expensive in a disk based system and so is only provided as an option at much reduced efficiency. With NVRAM as the media of storage, providing durability of data operations is no longer very expensive. Further, many data intensive applications need to implement transactions that provide ACID [20] properties to a sequence of operations. Since standard file systems do not provide transaction facilities, such applications either implement restricted versions of transactions [10], [2], or incur significant overheads to implement full ACID transactions [6]. [15] describes a system which provides ACID properties in an NVRAM file system. It however, requires non-volatile cache memory too. We have designed a file system that provides a restricted form of a transaction: operations between the open and the close of file automatically form a transaction with atomicity and durability guarantees. This feature can be used to build ACID transactions spanning multiple files, providing efficient implementation of RDBMS, NoSQL, and other data-intensive applications on NVRAM.

Our system contains the following novel features:

- It is designed on the premise that many applications will require support for atomic and durable operations, and so support for these should be provided at the file system level.
- We provide atomicity of file operations between an open and a close of a file, with only a successful close making all changes to the file permanent. We call a sequence of operations on a file starting with an open and ending with a close, a *transaction.* To support multi-file transactions, we plan to provide a new system call to close multiple files together. Storing rows of relations in separate files can then provide an efficient transaction implementation.
- We provide durability of changes to files at the close of a file. These features are over and above the consistency guarantees existing file systems on NVRAM provide.
- We use standard instructions, `clwb`, `sfence`, and `movnti` to implement the above features [8]. Further, since we have implemented on DRAM only, we do a read after write of the "last" change to ensure completion of changes. We are unable to use the "flush WPQ" feature of new Intel architectures [18].
- We show through an implementation and by comparison with another system NOVA [13], that the inclusion of atomicity and durability incurs acceptable loss in performance, which we assert will more than make up the overheads applications will otherwise have to incur if they need these features.

## 2. Related work

BPFS [3] is a file system on NVRAM. It uses the `movnti` instruction to make changes to a file system structure and thus ensures consistency of the file system at all times. Essentially, the last operation in a series of writes is a change to a pointer (of 64 bits). But the write to a file and updating its inode with the modification time cannot be done atomically. To order writes, they consider the use of `clflush` and `sfence` instructions to be inefficient to meet their requirements, and so they introduce two new hardware additions, one to ensure crash resistance of atomic writes using instructions like `movnti` (by adding capacitors inside the memory controller – Intel's ADR scheme now provides this), and another to ensure ordering (an epoch barrier instruction) which combines the operations of `clflushes` and `mfences`.

In PMFS [5], to achieve durability, a new hardware instruction (*pm_wbarrier*) is proposed. They also assume an optimised version of `clflush`. Their file system design is similar to BPFS using trees for metadata and using copy-on-write to ensure consistent updates through use of `movnti` instructions to update pointers. But they store meta-data in persistent memory and use a combination of in-place updates and an undo log for metadata updates. Data is accessed by user programs by memory mapping relevant PM areas to user address space. While this reduces one copy for data operations, implementation is complex, and introduces concerns of violation of protection domains. In the implementation described in the paper, PMFS only guarantees that the data becomes durable before the associated metadata does.

[9] evaluates the performance of standard Linux file systems and PMFS on NVRAM, and concludes that even standard file systems can be made efficient on NVRAM by proper tuning and configuration.

HiKV [12] is a key-value store implemented in NVRAM with a Hash Index in NVRAM and a B+ tree index stored in RAM to improve performance. Consistency of data in NVRAM is ensured by using 16 byte atomic writes using the `cmpxchg16b` instruction (with LOCK prefix). Durability is achieved by using `cflush` and `sfence` instructions. The focus on their paper is on the index design as they consider small key-values: 16 bytes of a key, and 500 bytes of value. They do not address the issue of operation completion.

DAX [17] support for File Systems in Linux is to enable implementation of file systems on NVRAM. DAX removes the Page Cache from such file systems, preventing a copy of data from the file system to RAM, and allowing direct access of the data from NVRAM. The support enables existing file systems such as EXT4 to be implemented on NVRAM. However, DAX does not provide any support to ensure consistency and durability of writes to NVRAM.

NOVA [13] (NOVA-FORTIS [14] is the system with reliability features included) is a file system implemented in PM. Each inode in the system is stored as a log. For each file and each directory, a DRAM based radix tree structure is maintained to provide fast access into its log. There is also a free block list kept in DRAM. These DRAM structures are not written to PM except when the file system is unmounted. Their design therefore reduces the amount of writes of meta-data to PM during run-time (keeping only the information in logs) at the cost of more complex recovery in case of a failure (essentially an "fsck" has to be done) and also more time to find attributes of files since they are stored in a log structure. They use copy-on-write to change data blocks, but they do not flush caches after every write to data blocks. Not doing this weakens consistency of data. Meta-data consistency is maintained by using `clwb` and `movnti` instructions. To cater to multiple CPUs, they have separate structures on a per-CPU basis.

SOUPfs [4] is a POSIX-compliant file system for NVRAMs. Since POSIX compliance does not impose deadlines for changes to be made durable, their system makes metadata persistent through background processes. In the critical path of file I/O no flushing of caches is done and so operation speed is improved. If there is a cache, there is always a consistent version available in persistent memory.

In [1], the authors consider alternative designs of a database system on NVRAM. They consider the use of a memory allocator library [11, 18], where durable operations are provided, and also the use of a standard POSIX file system with a "write behind log" for metadata. Being a tutorial, the paper does not describe any implementation, but it highlights some of the concerns of a database implementer, and which this paper seeks to address.

## 3. Design and Implementation

We have designed and implemented an ext3-like file system below the Linux VFS interface, for hosting in NVRAM. As hardware with NVRAM was not available, we designated a portion of RAM as persistent memory and carried out our implementation. The underlying persistent memory is partitioned into fixed size blocks of 4 KB. Like a traditional system, each file is identified by an *inode*. The inode structure is composed of only four fields: *i_blocks*, which represents the number of blocks being used by this object, a pointer *i_block* which points to a tree of pointers to data blocks (the tree grows as the file grows), *i_size* which represents the actual size of the file and *type* representing the type of the file: directory, file, symbolic link. We have not implemented ownership, access control etc.

Directories are files as in ext3, and their structure is also the same as that in ext3. The file system design is simple because all the metadata and code ensuring contiguity of related data in a file system, are no longer required.

| TS | BS | BB | IB | FB Map | FI Map | I Table | Log | Data Blocks |
|---|---|---|---|---|---|---|---|---|

**Figure 1: File System Layout**

The file system layout is given in Figure 1. Here the first 4 bytes (**TS**) represents the total size of the NVM in KB. The $5^{th}$ (**BS**), $6^{th}$ (**BB**) and $7^{th}$ (**IB**) bytes represent the block size in KB, number of blocks for the free block bitmap and number of blocks for the inode bitmap respectively. The free block bitmap (**FB Map**) is of size BS*1024*BB bytes which starts from the $8^{th}$ byte. The next BS*1024*IB bytes is the free inode bitmap (**FI Map**) and next to it, BS*1024*IB*8*sizeof (inode) bytes represents the inode table (**I Table**). The next K (configurable) blocks contains the log, and the remaining portion of the NVM is used as data blocks.

### 3.1. Metadata Redo Log

Due to durability and consistency issues, metadata and data cannot be written in place. A copy of each metadata is made to act as a buffer and changes are made to the buffer first. The buffer is implemented in DRAM to take care of cases where NVRAM access may be slower than that of DRAM. Changes made to metadata are also recorded in a write-ahead, redo log which is stored in NVRAM. For logging the metadata changes, we have implemented a very simple log structure which allows us to only log the metadata changes but not an entire metadata block. Space is pre-allocated for the log in the file system, and besides space for the log entries, there are two pointers, *start* and *end* which are stored along with the log in PM. The log space is used as a circular array with *start* and *end* denoting the start and end of the active log. New entries are always appended using *end*. On a failure, the recovery process will clear the log. But if there is no failure or a system shut down for a long time, the log may grow very long and space allotted to it may get exhausted, so, when the length of the log exceeds a pre-defined threshold value, the log gets cleared except for the currently executing transactions. Each log entry is 16 bytes long and is written into with two `movnti` instructions. *End* is updated using a `movnti` instruction, only after 16 bytes are written, and then an `sfence` instruction is issued. This ensures that on a crash, partial log entries cannot exist. This thus ensures an "all or nothing" update of the log. The first byte of a log entry defines the nature of the entry. The different entries of a log are shown in Table I. Depending on the *Entry Type,* entries *Data1, Data2,* and *Data3* may be present or the field may be blank. Every entry is identified with a transaction with a *Transaction No.* An entry contains an offset (previous pointer) to the previous entry made for this transaction. Entry types are : *set/reset inode bit map; set/reset free block bitmap; update inode logical block address; update inode i_size; update inode i_blocks; begin transaction; commit transaction; end transaction*. The only type with three parameters is *update inode logical block address*: Data1=inode number, Data2=logical block number, Data3: data block address.

When a transaction begins, a variable in the kernel keeps the index of the last entry for that transaction in the log. This is used when appending a new entry and the variable is updated with the index of the current entry.

| *Entry type* | *Data 1* | *Data 2* | *Data 3* |
|---|---|---|---|
| *transaction no* | *offset to prev entry of same transaction* | | |

**Table I Log Entry Format**

### 3.2. Data Blocks

We perform a copy-on-write whenever there is a write request to a data block. If a part of a block is to be written into, the block is copied to a free block and updates take place there. If a full block is to the written, then the write takes place on the new, free block. An update of the inode makes the changes permanent. In Figure 2 the sequence of operations for a low level write function, WRITE (*inumber, buffer, size, offset*), is given. *Buffer* is a pointer to user memory containing the data to be written to the file with inode *inumber*, *size* is the number of bytes to write, and *offset* is the offset in the file to write to. For ease of exposition, it is assumed that the write is to a single whole block or to part of one block only. The case for larger writes follows a similar pattern

```
Write(inumber, buffer, size, offset)

1.  Copy inumber from NVRAM to RAM if it
    is not already in RAM.
2.  Determine the logical block number I
    from offset and obtain the starting
    block S to which write has to take
    place from 'i_block' of inumber.
3.  Find a new free block F from the RAM
    copy of the free block bitmap (fbb)
4.  If partial modification of a block is
    to be done, Copy a block of data from
    S to F
5.  Write size bytes from buffer to F
6.  Flush cache lines containing data
    written into, using clwb
7.  Execute the sfence instruction;
8.  Set F^th bit of RAM fbb;
9.  Write  entry to log: "set F^th of fbb"
10. If partial modification
        Write entry to log: "reset S^th bit
```

```
   of fbb"
11. Update the RAM copy of the inode of
    inumber:
    (Block I pointer, i_size if required,
    i_blocks if required;)
12. Write entry to log: "block I = F of
    Inumber"
13. Write to log: "i_size of Inumber"
14. Write to log: "i_blocks of Inumber"
//log writes don't need clwb as movnti
instructions are being used
//write entry to log includes a sfence
instruction at the end.
```

**Figure 2: Write function**

### 3.3. Atomicity and Durability

In order to provide atomicity and durability, we need to define a point of durability for file operations. Under Linux semantics, there are no guarantees of durability of file writes since writes may be only to the buffer / page cache. A "flush" operation is required if durability is required. Even with this, durability support is weak, since there is no acknowledgement of storage completion in NVRAM. As we assume that the file system will be used to provide durable writes, we need to define a point of durability, which we define to be the closing of a file. On a successful close, the changes to the file is also made permanent, thus providing atomicity of operations between an open and a close of a file. Consistency and Isolation, if required, have to be provided by applications. Operations other than reads and writes are individually treated as atomic and durable.

### 3.4. Implementing Durability

As can be seen in Figure 2, after data is copied to block f, the cache lines are flushed. This is done to ensure that before making log entries for updating the inode of the file concerned (the log is in NVRAM), data written to block f is flushed. The `sfence` following the flush is to ensure that the log entry writes using `movnti` instructions take place after the data is in NVRAM. The `clwb` instruction is of the form `clwb` *m8*. The cache line containing the memory address *m8* is flushed to main memory. The cache line size is 64 bytes. So, to flush a 4KB data block whose starting address is A, we have to issue 64 `clwb` instructions of the form `clwb` *A*, `clwb` $A+2^6$, `clwb` $A+2^6+2^6$, *and so on (64 X 64 = 4KB).*

When a file is closed, the situation is as follows: all data writes have been made to NVRAM and cache flushes issued for these writes. All changes to metadata as a consequence of the file operations have been made in the RAM copy of the concerned metadata (except for release of resources). All metadata change operations are written into the log. Now, a *commit* transaction record is written to the log and then the changes recorded in the log for this transaction are made in the actual metadata locations in NVRAM. When all such changes have been made, an *end transaction* record is written into the log. If a metadata change involves a *reset* of either the free block bit map or a free inode bitmap, then these changes are not made in the RAM version of these bitmaps, as mentioned as a comment in the code in Figure 2. Suppose transaction A deletes a block x from the file it is operating on. The $x^{th}$ bit of the free block bit map has to be reset. But this should be done only when the transaction commits as otherwise there can be problems. If a subsequent transaction B requires a free block, it may find x and use it. Later, transaction B commits, but before A can commit, a failure occurs. So no changes are made to the inode of transaction A and one of its entries is still pointing to x. But x has been reused by B. On the other hand, if transaction A requires a free block and acquires block y, it must set the $y^{th}$ bit of the free block bit map as otherwise transaction may also use block y, finding it free. During recovery, if it is found that transaction A has not committed, while nothing needs to be done in the metadata in NVRAM, the setting of bit y in the free block bit map in the RAM copy must be *undone.* The sequence of operations on a commit is:

```
1. Write Commit Record in log;
2. Execute sfence
3. Read the last record of the commit
   record and compare with the value
   written (stored in a CPU register); If
   the value read matches, commit record
   (and all previous records) are durable
   in NVRAM.
4. Follow the back pointers in the
   transaction records, and for each
   record, make necessary changes in the
   metadata in the file system in NVRAM.
5. If a change is a reset to a bitmap,
   make that change to the RAM version
   too.
6. Flush all cache lines involving the
   metadata changes
7. Execute sfence
8. Write "end transaction" record in the
   log.
```

We must ensure that the commit record is durable before we can make changes to the metadata in NVRAM. As already discussed, the only way to ensure this is to reread the data already written into NVRAM. Since we are using `movnti` to write into the log, and since `movnti` bypasses the cache hierarchy, there is no need to flush any cache lines. The `sfence` is required though, to ensure that metadata changes are made to NVRAM after the commit record actually reaches NVRAM.

### 3.5. Recovery

If a crash occurs, we redo the entire log for all transactions that have a *commit transaction* but not an *end transaction* entry in the log. Changes are written to the actual locations of the metadata in NVRAM. The RAM version of the metadata has in any case got wiped out due to the crash. Because of the `sfence` instruction between data block writes and the log write, if we find the log write in NVRAM after a crash, it means that the data writes have also reached the NVRAM (`sfence` ensures that previous writes to memory take place before subsequent writes). If there are log entries for a transaction whose *commit transaction* entry is not there in the log, then that transaction is assumed to have been aborted. Since changes to metadata due to this transaction (which are in the log) are not redone, the effects of the transaction are not persisted. The data blocks that were written into may have been written into NVRAM, but they remain free in the free block bitmap and so the contents are not used. Once a transaction's changes in the log have been made permanent in NVRAM metadata, an *end transaction* entry has to be written. In the next round, when *end transaction* is found, we are sure changes are permanent and so the entries of this transaction can be removed. The basic scheme is the similar to the ARIES system [7]. However, since the log is only redo and there is no undo, there is no need to write "compensatory log entries" as described in ARIES. A failure taking placed during a recovery has no impact, in that in the second recovery, the entire process will be repeated. Since all changes to metadata are idempotent, repeated changes to the same metadata can be made.

### 4. Experiments

Due to shortage of space we describe only one of the experiments. Experiments were carried out in a 6 core Xeon system with 32GB of memory. 4GB was assumed to be NVRAM and set up accordingly in the kernel. Comparison was done with the NOVA file system which was also installed in the same system. We used Linux kernel version 4.13.0. We ran two benchmarks Fio[16] and Filebench[19] on NOVA and on our system. The benchmark workload characteristics are given in Table II. For DurableFS, there were two versions: 1) one the atomic and durable version, and 2) the second in which data writes were not flushed from cache (NOVA too does not do this). Version 2 was compared with NOVA to assess the efficiency of our implementation. Comparison between version 1 and 2 gave us the degradation in performance due to flushing all data writes.

Figure 3 gives the results of our experiment. Each experiment was run 10 times and the average result was taken. As can be seen, NOVA is about 2% faster than version 2 of DurableFS in Fio and Filebench (fileserver) workload. On the other hand NOVA is about 1% slower than version 2 of DurableFS in the Filebench (webserver) workload, which is a read intensive workload (See the R/W ratio in Table II). These differences are not significant. Comparison between version 1 and version 2 shows that there is a degradation of performance of about 7%. We feel this is reasonable, given the extra functionality being provided to applications.

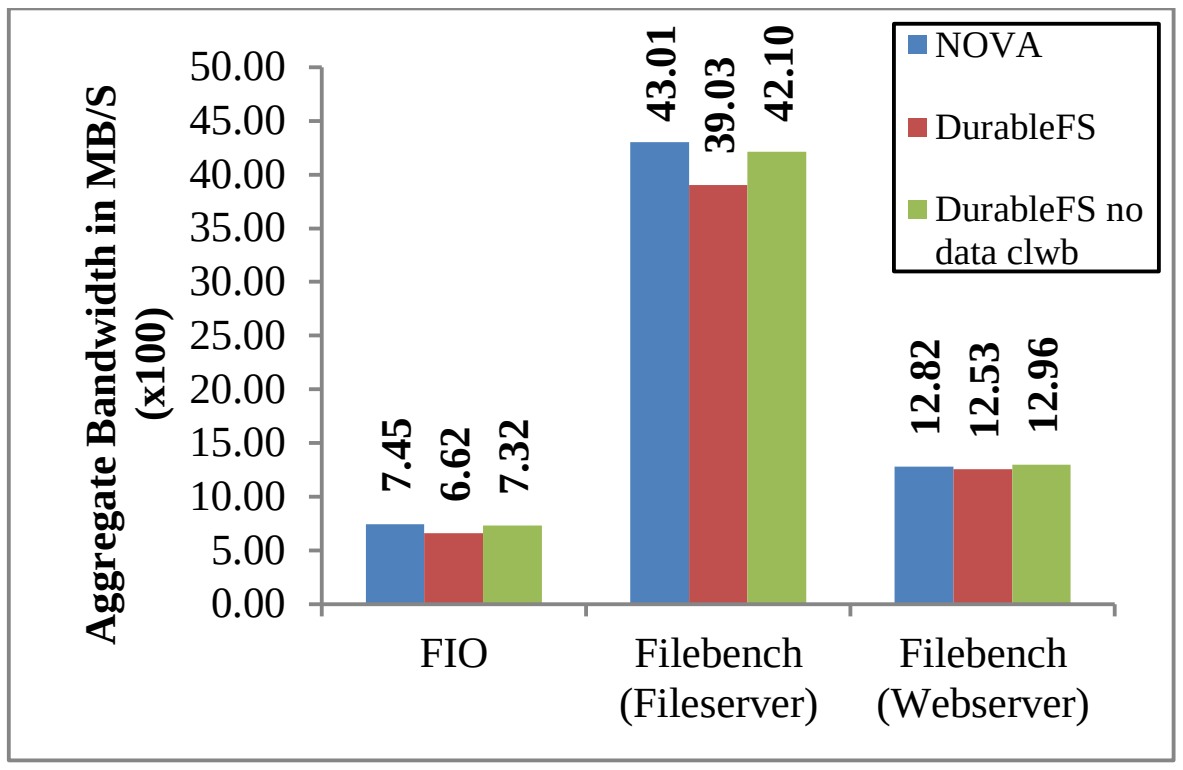

**Figure 3 Benchmark Results**

| Workload | File Size | I/O Size | Threads | R/W Ratio | No of files |
|---|---|---|---|---|---|
| Fio | 256M | 4K | 10 | 1:1 | 10 |
| Filebench (Fileserver) | 128K | 4K | 10 | 1:2 | 1K |
| Filebench (Webserver) | 64K | 4K | 10 | 10:1 | 1K |

**Table II Benchmark workload characteristics**

### 5. Conclusion

We have implemented a file system for persistent memory (NVRAM) on the memory bus as a Linux kernel module. The store uses copy-on-write for writing into data blocks, and it uses a redo log to record changes to metadata while making the changes to versions of the metadata buffered in RAM. Durability of data in NVRAM is ensured by flushing data in caches of the CPU by using `clwb` and `sfence` instructions. Consistent updates are made using `movnti` instructions. The point of durability is the closing of a file, and it is also the point of atomicity. Experiments showed that the overhead due to the flushing of caches to ensure consistency of data resulted in only 7% degradation in aggregate bandwidth on an average. This demonstrates that NVRAM can be used efficiently and durably as a store.

We propose to convert the file system to a key-value store and to build a distributed key-value store on a clustered system with NVRAM at each node.